\documentclass{osa-article}

\journal{oe}

\articletype{Research Article}

\usepackage{lineno}
\usepackage{booktabs}
\usepackage{tabularx}
\usepackage{hyperref}
\hypersetup{colorlinks=true, linkcolor=blue, citecolor=blue, urlcolor=blue, unicode=false}

\begin{document}

\title{An efficient low-density grating setup for monochromatization of XUV ultrafast light sources}

\author{Qinda Guo,\authormark{1} Maciej Dendzik,\authormark{1} Magnus H. Berntsen,\authormark{1} Antonija  Grubi\v{s}i\'{c}-\v{C}abo,\authormark{1,2} Cong Li,\authormark{1} Wanyu Chen,\authormark{1} Yang Wang,\authormark{1} and Oscar Tjernberg\authormark{1,*}}

\address{\authormark{1}Department of Applied Physics, KTH Royal Institute of Technology, Hannes Alfv\'{e}ns v\"{a}g 12, 114 19 Stockholm, Sweden\\}
\address{\authormark{2}Zernike Institute for Advanced Materials, University of Groningen, 9747 AG Groningen, The Netherlands\\}

\email{\authormark{*}oscar@kth.se} 



\begin{abstract}

Ultrafast light sources have become an indispensable tool to access and understand transient phenomenon in material science. However, a simple and easy-to-implement method for harmonic selection, with high transmission efficiency and pulse duration conservation, is still a challenge. Here we showcase and compare two approaches for selecting the desired harmonic from a high harmonic generation source while achieving the above goals. The first approach is the combination of extreme ultraviolet spherical mirrors with transmission filters and the second approach uses a normal-incidence spherical grating. Both solutions target time- and angle-resolved photoemission spectroscopy with photon energies in the 10-20 eV range but are relevant for other experimental techniques as well. The two approaches for harmonic selection are characterized in terms of focusing quality, efficiency, and temporal broadening. It is demonstrated that a focusing grating is able to provide much higher transmission as compared to the  mirror+filter approach (3.3 times higher for 10.8~eV and 12.9 times higher for 18.1~eV), with only a slight temporal broadening (6.8\% increase) and a somewhat larger spot size ($\sim$30\% increase). Overall, our study establishes an experimental perspective on the trade-off between a single grating normal incidence monochromator design and the use of filters. As such, it provides a basis for selecting the most appropriate approach in various fields where an easy-to-implement harmonic selection from high harmonic generation is needed.
\end{abstract}

\section{Introduction}
Extreme ultraviolet (XUV) radiation with a narrow bandwidth is one of the key ingredients for many high-resolution spectroscopic techniques used in research today. Beyond this, XUV light also has considerable application value for the industry, for example in lithographic patterning adopted in semiconductor manufacturing processes. With the advent of femtosecond high-power lasers and the progress of high harmonic generation (HHG) techniques, table-top XUV light sources with pulse duration in the femtosecond range have become a reality. These sources offer a more compact and affordable alternative compared to the synchrotron radiation facilities, or free-electron lasers, while simultaneously supporting spectroscopic, imaging or scattering techniques with high temporal resolution. Presently, HHG has become a viable approach to generate bright and ultrafast XUV-light sources, but monochromatization of the generated radiation still remains a challenge -- in particular from the point of view of maintaining a high photon transmission through the monochromator, while simultaneously minimizing time broadening of the light pulses, as well as keeping the optical design robust and compact. 

HHG is a well-developed technique used to convert long-wavelength laser light into radiation with shorter wavelengths, typically in the XUV or soft X-ray regime\cite{LaserHHG_KepteynMurnanePRL1996, HHG_CoherentSoftXrays_Murnane_PRL1997, HHG_CoherentXrays_WaterWindow_Krausz_Science1997}. The HHG process can be understood conceptually by the three step model\cite{HHG_Ne_Ar_LHullier_PRA1993,HHG_threeStepModel}: firstly the electrons experience the ionization tunneling, followed by a propagation in the electric field from the driving light, and finally recombination with the ionic nuclei, yielding high energy and coherent photons during the process. The HHG spectrum consists of a series of discrete harmonics spaced by twice the driving frequency, 2$\omega$, up to a maximum photon energy determined by the single atom cut-off relation, $h\nu_\mathrm{cut-off}= I_\mathrm{p}+3.17U_\mathrm{p}$, where the $h$ is Planck's constant, $I_\mathrm{p}$ is the ionization potential and $U_\mathrm{p}$ is the ponderomotive energy of the electron. This relation implies that the energy cut-off can be increased by using a longer driving wavelength, as it scales with the relation $U_\mathrm{p} \propto I_\mathrm{L} \lambda_\mathrm{L}^2$, where $I_\mathrm{L}$ is the intensity of the driving laser and $\lambda_\mathrm{L}$ is the wavelength. On the other hand, the HHG efficiency scales proportional to $\lambda_\mathrm{L}^{-(5-6)}$\cite{tate2007scaling}. This inherent trade-off means that, in practice, one needs to choose suitable driving conditions based on the light source requirement. In our case, to maximize the photon flux at photon energies up to 25 eV, a cascaded scheme\cite{comby2019cascaded}, using the third-harmonic of the driving laser, is chosen for HHG in a gas target.

The monochromatization requires a selective response to different frequency components, and gratings are extensively adopted for this purpose. A typical monochromator layout is the X-ray Czerny-Turner (XCT) design\cite{Mono_CzernyTurner1930}, which is a single-grating based grazing incidence monochromator (GIM), that provides the most necessary elements needed for the collimation, diffraction and re-focusing. The line density is usually chosen based on the trade-off between efficiency and spectral resolution. With high line-density, very high resolving power and energy resolution have been achieved \cite{peatman1995exactly,borisenko2012one}. Another common GIM layout is a spherical grating based configuration, which serves as a compact layout\cite{peatman1995exactly,hoffmann2004undulator}. As the counterpart of GIM, the normal incidence monochromator (NIM) has also been employed in synchrotrons to access the low photon energy range and provide high energy resolution\cite{borisenko2012one}. Compared to GIM, NIM is quite advantageous in terms of compact layout, ease of alignment, and good XUV imaging properties.

Stepping into the ultrafast regime in the XUV range, monochromators based on diffractive optics exhibit intrinsic drawbacks, as the induced pulse-front tilt broadens the pulse length. Nevertheless, there have been designs targeting HHG \cite{poletto2004time,poletto2009time,ito2010spatiotemporal,frassetto2011single,igarashi2012pulse,grazioli2014citius,ojeda2016harmonium} and free-electron-laser (FEL) light sources \cite{heimann2011linac,frassetto2014time,fabris2019high}. One approach is to use a single-grating, but lowering the line density to reach the design requirements \cite{grazioli2014citius,poletto2014double,ojeda2016harmonium}, concerning the pulse length trade-off for a certain spectral linewidth.  The double-grating design is an alternative approach that permits complete pulse stretching compensation and ideally provides no extra temporal broadening \cite{poletto2004time,ito2010spatiotemporal,igarashi2012pulse,fabris2019high}. Pulses as short as 8~fs have been reported \cite{poletto2009time} using this design. However, such a configuration reduces the transmission and makes the design and alignment somewhat complicated. The pulse stretching induced by a single grating can, however, be reduced by using the off-plane-mount. In this case the diffraction occurs in the direction perpendicular to the direction of propagation thus minimizing the number of illuminated groves and thus decreasing the temporal broadening \cite{frassetto2011single,frassetto2014grating,poletto2014double}. 

\begin{figure}[htp]
\centering
\includegraphics[width=\textwidth]{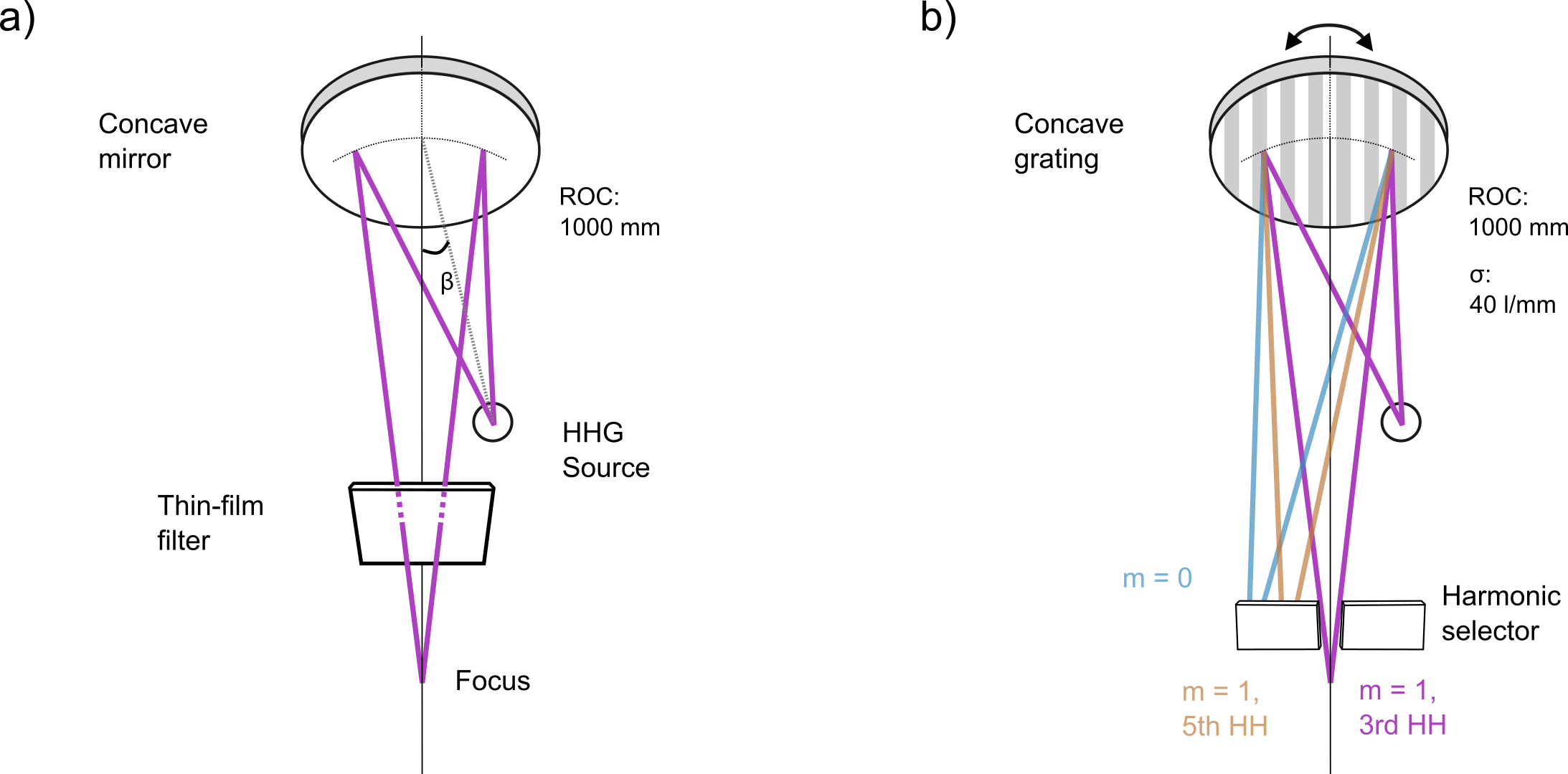}
\caption{\label{fig:nims}Schematic drawing of the two types of monochromatization for ultrafast applications. a) The focusing mirror plus the thin film filter. b) The focusing grating. ROC stands for the radius of curvature, m is the diffraction order, $\sigma$ is the line density and $\beta$ is the beam incidence angle. HH stands for high harmonic.}
\end{figure}

In this work, we compare the performance of two types of NIM configurations, as illustrated in Fig.~\ref{fig:nims}. The mirror+filter approach and the spherical grating monochromator are compared in terms of their focusing properties, efficiencies and the introduction of temporal broadening of the HHG pulses. The ultrafast light source is part of an experimental setup for time- and angle-resolved photoemission spectroscopy (tr-ARPES)~\cite{guo2022narrow}. For the comparison, we designed a spherical grating NIM specially for 10.8~eV and 18.1~eV photon energies. In this particular case, it is found that the mirror+filter design results in a slightly better focus, 100~{\textmu}m spot size, while the focusing grating focus is about 30\% larger ($\sim$33~{\textmu}m). The larger spot size could be due to fabrication imperfections of the grating profiles or overcoating. On the other hand, in terms of transmission efficiency, the focusing grating is remarkably effective for both photon energies, resulting in 3.3~times higher efficiency for 10.8~eV and 12.9 times higher efficiency for 18.1~eV as compared to the mirror+filter solution. Consequently, this enabled us to achieve a photon flux on the level of 10$^{11}$ photons/s. As for the temporal response, the mirror+filter combination, which is used as a reference, is assumed to provide negligible temporal broadening. Pump-probe measurements on graphene show an overall temporal resolution of 205.3$\pm$7.6~fs for mirror+filter and 219.5$\pm$1.9~fs for the spherical grating design. By excluding the temporal contribution from the pump, this corresponds to a temporal broadening of 72.4$\pm$27.4~fs. In Table 1, the various performance parameters are listed, to compare the applicability of the grating and filter designs for a specific experimental design.

\section{Monochromatization of an ultrafast light souce}
For monochromatization of ultrafast light sources, the mirror+filter configuration is preferred in terms of time resolution as it can eliminate the temporal broadening issue, providing a significant advantage compared to monochromators based on gratings. In our setup, the mirror+filter configuration is comprised of a concave normal incidence mirror, and a thin film filter that is used to further suppress unwanted harmonics. The mirror coating method is critical for the harmonic selection, but it is difficult to completely eliminate unwanted harmonics with the coating alone. For this reason, the thin film filter with an appropriate transmission window is necessary. The drawback of this approach is that the performance and working wavelength range are primarily determined by the available thin film filters, and they do not always fit the experimental demands. Since thin film filters need to be as thin as hundred nanometers to maintain reasonable transmission, they are delicate and can easily be damaged mechanically or by the beam. Oxidation and contamination are other issues often experienced in the use of thin film filters.

Here we propose a novel and efficient grating-based NIM that has been designed, fabricated and experimentally compared with the mirror+filter solution. It is in principle a spherical focusing grating, which serves as a single optical element-based NIM, with a low line density of 40~l/mm. The main design criteria is to have as low line density as possible while still being able to separate the harmonics geometrically at the sample position. In the present context, the low resolving power caused by the low line density is not an issue, as the spectral resolution is intrinsically determined by the HHG conditions. The purpose of using the lowest possible line density is to mitigate the pulse-front tilt to a negligible level. Note that the value of 40~l/mm adopted here, is dependent on the particular geometry of the setup and the trade-off between pulse stretching and focal properties. In the present comparison, the two compared solutions have similar geometry and focusing properties so that the performance can be compared in a relatively straightforward manner. Depending on the availability of filters and coatings the two solutions are also more or less difficult to implement for various wavelengths, in the present case we target photon energies ($\sim$10-20~eV). For the spherical grating efficiency, the simulation with a groove depth of about $\sim$ 17~nm, together with the 80~nm SiC coating, shows a 15\% overall efficiency centered at 18.1~eV photon energy, and overall, above 10\% efficiency in the range from 10 to 20~eV, as presented in Fig. \ref{fig:sim}(c).

\begin{table}[]
\caption{\label{tab:criteria} Criteria for two configurations of ultrafast XUV pulse monochromatization.}
\resizebox{\textwidth}{!}{
\begin{tabular}{lcc}
\hline
 & Mirror+filter & Focusing grating \\ \hline \hline
Pulse-front tilt & Negligible & Below 100 fs for $h\nu>10$\,eV$^\mathrm{\dagger}$ \\ \hline
Spectral broadening & Negligible & Negligible \\ \hline
Efficiency & \begin{tabular}[c]{@{}c@{}}Moderate. Limited by reflectivity\\ of the mirror and transmission\\ through the thin filter\end{tabular} & \begin{tabular}[c]{@{}c@{}}High. Limited by reflectivity of\\ grating, which is set by the groove \\ depth, groove-to-land ratio\\ and coating material\end{tabular} \\ \hline
Focusing & \begin{tabular}[c]{@{}c@{}}Determined by source size\\ and spherical aberration\\ from focusing mirror\end{tabular} & \begin{tabular}[c]{@{}c@{}}Broadened compared to spherical\\ mirror. Mainly limited by\\ the fabrication errors\end{tabular} \\ \hline
Costs & \begin{tabular}[c]{@{}c@{}}High. Requires multiple\\ band pass mirrors as\\ well as thin film filters\end{tabular} & \begin{tabular}[c]{@{}c@{}}Moderate. Single optical element\\ low-line-density grating produced\\ by standard nanofabrication\\ processes\end{tabular}    \\ \hline \hline

\end{tabular}
} 
\footnotesize{$^\mathrm{\dagger}$ This is for the experimental design presented in this article. The temporal broadening can be further decreased by lowering the number of illuminated grooves through lower line density, or smaller beam spot size.}
\end{table}

\section{Mirror+filter configuration}
For the direct comparison of the filter and grating-based monochromator solutions, we focus on two photon energies, 10.8~eV and 18.1~eV, generated by the HHG setup. For the mirror+filter configuration, and for a photon energy of 10.8~eV, a concave MgF$_2$-coated Al mirror with a radius of curvature (ROC) of 1000~mm is used to focus the diverging high harmonic radiation.
The filter in this case is a thin (500~{\textmu}m) LiF (Eksma Optics) window that blocks photon energies higher than 11~eV.
For 18.1~eV photons, the previous mirror and filter were replaced with a concave silicon mirror substrate (ROC=1000~mm) coated with SiC and a 150~nm thick Sn filter (Lebow), respectively. The Sn filter provides a band-pass window for photon energies between 17~eV and 24~eV. For more information on the band-pass windows for different thin film filters, see Supplementary Material Fig. S4.

\section{Focusing grating configuration}
With only a few changes, the experimental setup of the mirror+filter monochromator described above can be turned into a grating-based NIM. The key is to replace the spherical focusing mirror with a spherical grating with the same ROC and instead of thin film filters use a slit to select the desired diffraction order from the grating. In the following section, we elaborate in detail on the focusing grating part, in particular regarding the design principles, and the fabrication recipes.

\subsection{Design principles}
To design a NIM using a focusing grating, and to match the specific requirements regarding high efficiency and temporal resolution, several constraints need to be considered. Compared to the mirror+filter solution, one major challenge for designing the focusing grating is to limit temporal broadening of laser pulses. The pulse-front tilt experienced on the grating is proportional to the number of grooves that are illuminated, resulting in a correlation between the line density, $\sigma$, and the illuminated area, $A$, on the grating. The spot size $A$ is given by the HHG divergence and the focusing geometry of the setup and, therefore, in this work, it is regarded as a fixed parameter. The illuminated area on the grating can be modulated by an aperture at the cost of reduced photon flux. Keeping the line density $\sigma$ low can minimize time broadening of the pulses, but a sufficiently large $\sigma$ is required to spatially separate the different diffraction orders of the harmonics at the position of the focus. The groove depth, $d$, and the ratio of the groove and peak widths, $r$, determine the efficiency and the band-pass center of the grating. The choice of a coating, and its thickness, are also factors affecting the final efficiency. 

\begin{figure}[htp]
\centering
\includegraphics[width=\textwidth]{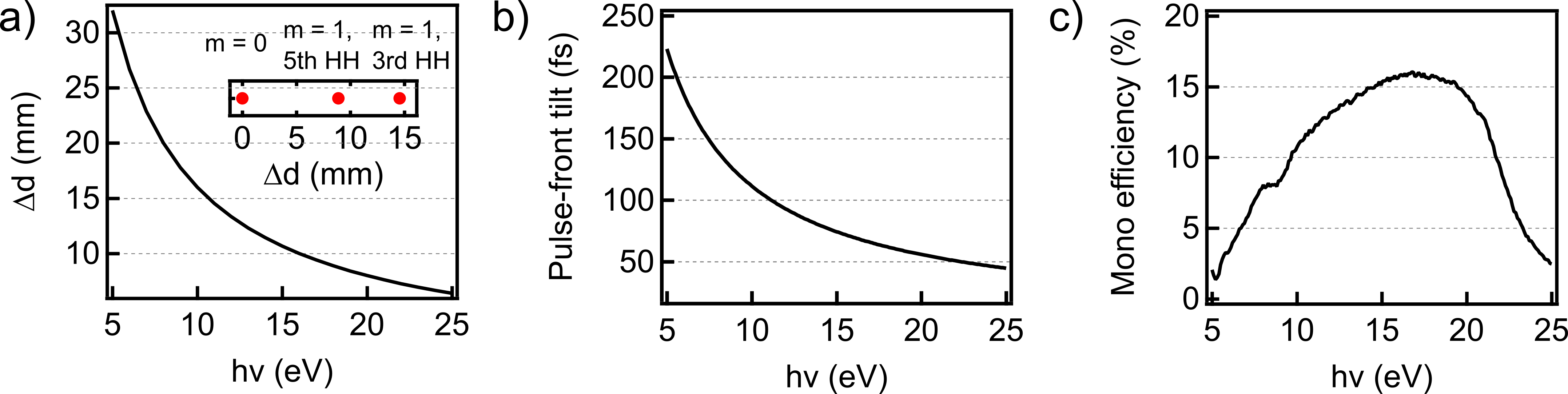}
\caption{\label{fig:sim} The simulation results for the spherical grating NIM based on the following principle parameters: 40~l/mm line density, 17~nm groove depth, 1:1 land to groove ratio, and 80~nm SiC coating. a) The distance of the first order higher harmonic spots with respect to the position of the zero diffraction order. b) The pulse-front tilt, and c) the overall efficiency, plotted as a function of the photon energy in range from 5 to 25~eV. The focal length is set to 3~m.}
\end{figure}

Figure~\ref{fig:nims}b) depicts the concept of the spherical grating-based monochromator that reflects and refocuses the diverging HHG radiation generated in the gas target, while simultaneously separating the harmonics into different diffraction spots. The desired harmonic can then be selected with a slit or an aperture. For a grating with $N$ grooves illuminated by the incident beam, the resolving power, $R$, is expressed as: 
\begin{equation} \label{eq:1}
R = \frac{\lambda}{\Delta \lambda} = |m|N = \frac{\Delta t}{\lambda}
\end{equation}
where $m$ is the diffraction order, $\Delta t$ is the total pulse-front tilt and $\lambda$ is the radiation wavelength. The expression implies a proportional relation between the resolving power and the induced time broadening for a grating-based monochromator. In our case the spectral linewidth is determined by the HHG conditions, so as a design principle, one should pursue a low resolving power for the minimum temporal broadening, which directly links to the line density $\sigma$. Another constraint for $\sigma$ is the wavelength spacing between the different harmonics, as presented in Figure~\ref{fig:sim}b), since there needs to be sufficient spatial separation for an aperture to select the harmonics.

Following the aforementioned design principles, a focusing grating is explicitly designed for a photon energy of 18.1~eV. The chosen line density $\sigma$ is 40~l/mm. With the distance from the grating to the sample in the photoemission chamber being 3~m at the sample position, the first order diffraction spots for 18.1~eV and 10.8~eV are positioned 8~mm and 14~mm away from the zero-order diffraction maximum, respectively. This separation is sufficient so that an aperture can be used to select the desired harmonics and block the remaining undesired diffraction spots. The 17~nm groove depth $d$ and the 1:1 ratio value of $r$, provide optimal efficiency for the 18.1~eV photon energy. The grating pattern is etched into a silicon mirror substrate with a ROC of 1000~mm, and finally, a SiC coating layer with a thickness of minimum 30~nm was deposited. This minimum coating layer thickness yields the best performance according to calculations (see Supplement Material Section 1 and Figure S1.). Figure~\ref{fig:sim} presents calculations of key performance parameters for the focusing grating for the photon energy range between 5~eV and 25~eV, using the parameters presented above. Figure~\ref{fig:sim}a) presents the calculated space-separation of the diffracted harmonics. The distances between the zeroth order and the first diffraction order for specific photon energies, are plotted. The focal length is set to 3~m. The inset shows the results of separation for the HHG conditions and setup geometry in our case. Figure~\ref{fig:sim}b) shows the pulse-front tilt calculated based on Eq.~\ref{eq:1}, as a function of photon energy. The total efficiency of the grating is illustrated in Fig.~\ref{fig:sim}c).

\subsection{Fabrication recipe}
\begin{figure}[htp]
\centering
\includegraphics[width=\textwidth]{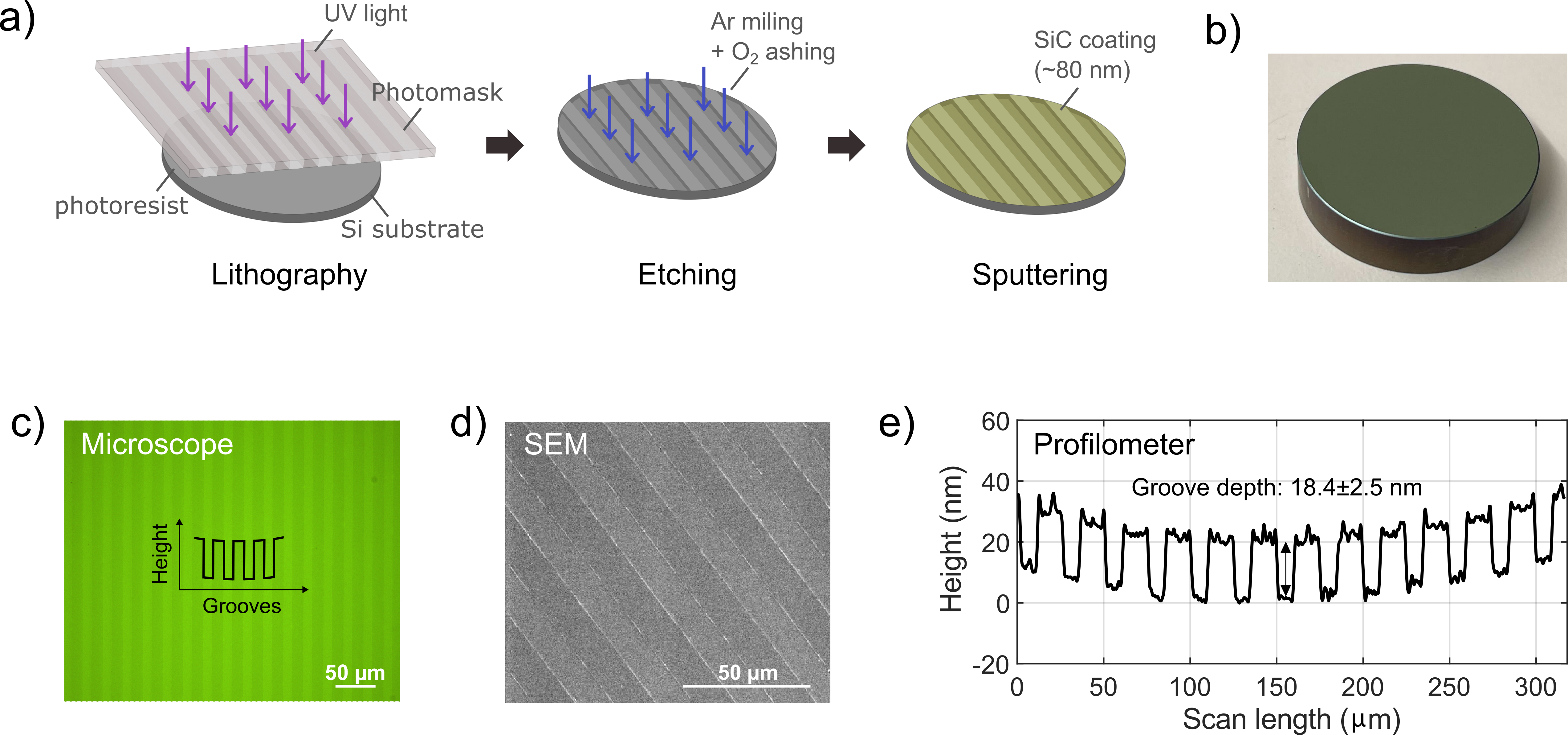}
\caption{\label{fig:fab}Fabrication of the spherical grating. a) Schematic drawing of the fabrication processes. b) The final piece after the fabrication. c) and d) present the optical microscope and SEM images taken on the grating shown in b), respectively. e) Profilometer characterization results showing an average groove depth of 18.4$\pm$2.5~nm.}
\end{figure}

The fabrication of the spherical grating was conducted in a clean-room using standard semiconductor processing steps such as lithography, chemical etching, and sputter deposition of the coating material. The different steps of the manufacturing process are schematically shown in Fig.~\ref{fig:fab}a). The plano-concave 1~inch silicon mirror substrate (EKSMA Optics), is first spin-coated with a photoresist (Microposit, S1813) on the concave surface. A photomask made of chromium is used in the exposure step with a high dose exposure (30~mW/cm$^2$ at 365~nm) from a helium lamp (Karl Suss MJB3). The exposure time was 120~seconds. The substrate was then transferred into a reactive-ion-etching (RIE) system (Oxford instruments) where argon plasma milling for $\sim$ 6~min was used to etch the grating pattern and reach the desired groove depth ($\sim$17~nm). The etching was followed by a soft ashing procedure with oxygen in order to remove the residual photoresist. As the final step, the substrate was coated with SiC in a thin-film deposition system (AJA Orion-8) to enhance the reflectivity in the XUV range. The fabricated piece, and the overall quality of the etched pattern, was then examined with an optical microscope and a scanning electron microscope (SEM; FEI Nova 200). The results are shown in Fig.~\ref{fig:fab}c) and d), respectively. The depth and the ratio of the grooves were measured with a stylus profilometer (KLA Tencor) and are shown in Fig.~\ref{fig:fab}e).

One major issue limiting the fabrication quality for such a grating is the nature of the curved surface, as this causes non-uniformity in terms of the groove depth $d$ and the ratio $r$. With a ROC of 1000~mm for the concave surface of the mirror substrate, there is a height difference of $\sim80$~{\textmu}m between the rim and the center of the mirror. This makes it challenging to achieve a homogeneous thickness of the photoresist layer, as well to have an even exposure of the photoresist since the curvature leads to the UV light being defocused towards the center compared to the rim of the mirror. Simulations (see Supplementary Material Section 2 and Figures S2 and S3) show that the value of the groove depth dominantly affects the band-pass center, but the efficiency is almost unchanged for a $\pm$2.5~nm tolerance of the depth. This requirement can be fulfilled in the fabrication used here. As Fig.~\ref{fig:fab}e) shows, the groove depth is about 18.4$\pm$2.5~nm which is within the range of optimal performance.

\section{Experimental setup}

\begin{figure}[htp]
\centering
\includegraphics[width=\textwidth]{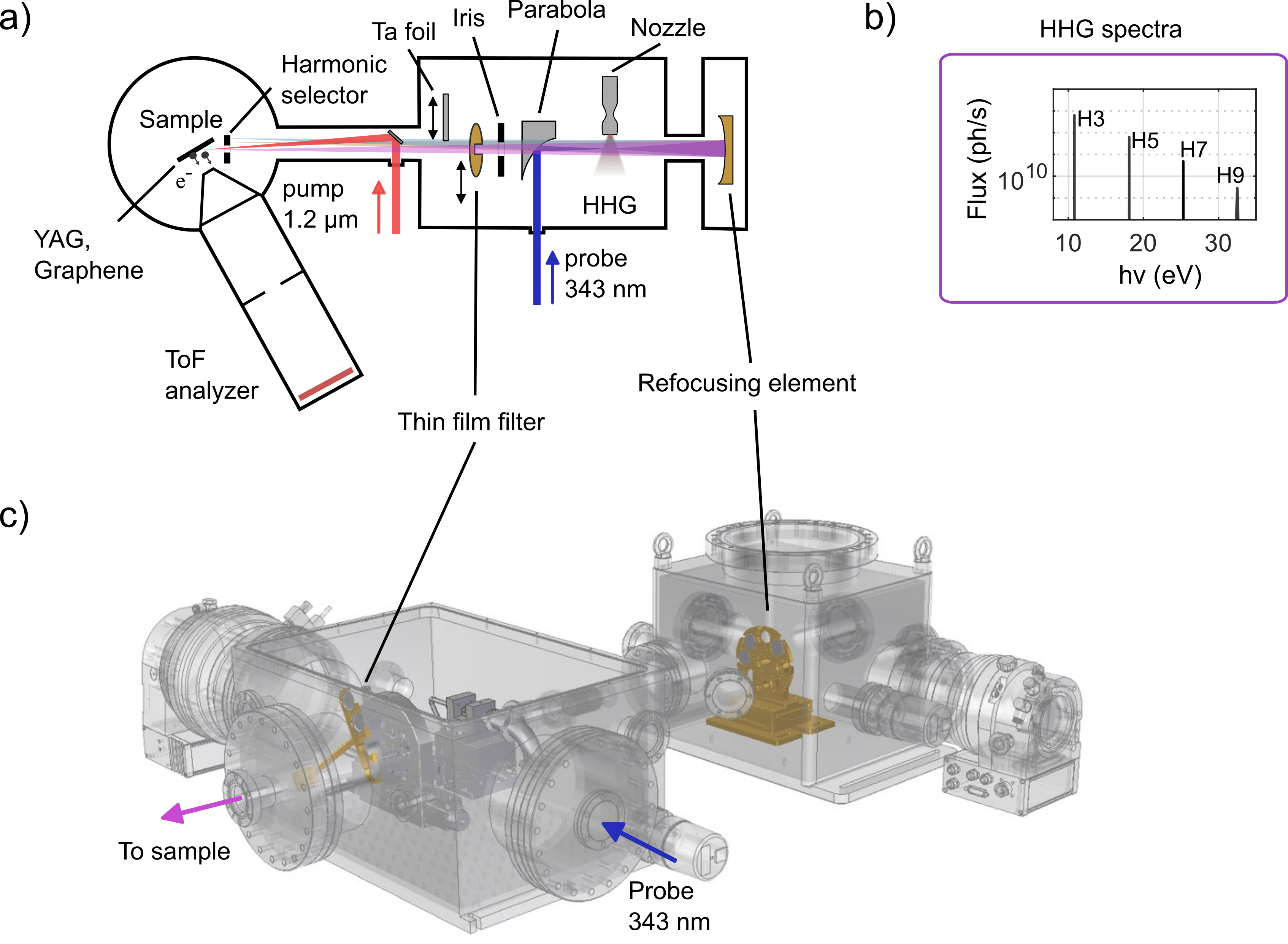}
\caption{\label{fig:setup}The experimental setup used for characterizing the monochromator. a) The schematic drawing of the beam path. b) HHG spectral flux on sample as a function of photon energy. c) Model of the HHG and refocusing chamber showing detailed mounting and placement of the optical elements for the example configuration. Mono-components are marked by golden color.}
\end{figure}

The performance of the two types of monochromatization is characterized using the HHG-based light source and time- and angle- resolved photoemission spectroscopy (tr-ARPES) setup described in Ref.~\citeonline{guo2022narrow}. This system yields ultrafast, few-hundred-femtosecond long XUV pulses. Figure~\ref{fig:setup} shows the schematic diagram of the setup, illustrating the specific optical layout. Briefly, the HHG light source consists of two Ytterbium-Doped Fiber Amplifiers (YDFA) lasers, one (Tangor, Amplitude Systems) for driving the HHG of the probe source, and another (Tangerine, Amplitude Systems) for driving an optical parametric amplifier (OPA) which provides the pump beam. Both lasers deliver pulses with a wavelength centered at 1030~nm, and are physically synchronized to share one common oscillator. For the pump line, the infrared (IR) beam has a duration of about 280~fs and is compressed from the OPA to $\sim$100~fs at an output wavelength of 1.2~{\textmu}m, which is used for the temporal response measurement shown in the present work. For the probe line, the fundamental light source has maximum pulse energy of $\sim$300~{\textmu}J at a 250~kHz repetition rate. The pulse duration is $\sim$461~fs. Prior to the HHG, the IR output is firstly tripled in frequency through a two stage up-conversion using non-linear optical crystals, to 343~nm, with an efficiency of about 30\%. The UV light is focused by an off-axis parabolic mirror into a gas jet for HHG process, which generates output in the XUV range from 10.8 eV to 32.5~eV. The HHG is conducted in the tight focusing geometry with argon gas as the medium. In this work, we focus on the 3rd ($\sim$10.8~eV) and 5th ($\sim$18.1~eV) harmonic. Following the HHG, a mirror wheel carrying the refocusing optics is placed, with the incidence angle of $\sim$1$^{\circ}$.

\section{Performance}
Three fundamental properties are characterized to evaluate the performance of the two types of monochromatization investigated in this work. These are: optical quality (focusing), efficiency, and temporal broadening. The experimental results for the mirror+filter based and the grating-based optical setups are presented, and compared, below.

\subsection{Optical quality and focusing}
\begin{figure}[htp]
\centering
\includegraphics[width=\textwidth]{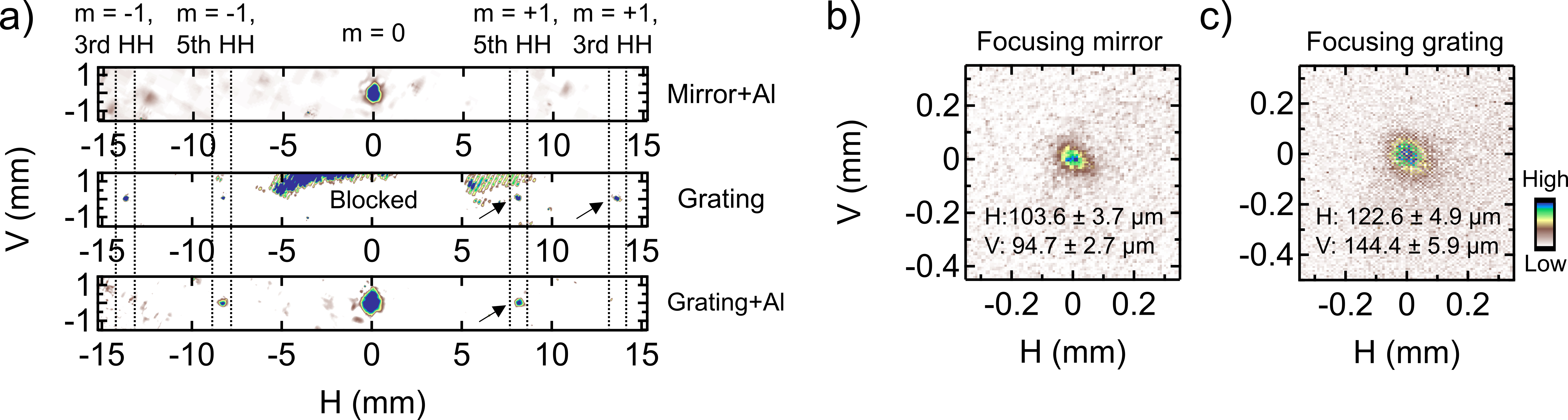}
\caption{\label{fig:spot}The analysis of the diffracted beam spots from the fluorescence screen. a) The observed spot pattern based on different combinations: mirror+filter (top), grating (middle) and grating/filter (bottom). The ability to separate the harmonics is clearly demonstrated. b) and c) depict the spot size measured with a YAG crystal for the mirror+filter and the focusing grating setups, respectively. Horizontal and vertical spot size is marked in the figures.}
\end{figure}

The focusing properties of both the simple spherical mirror and the grating are characterized by observing the light spots on a fluorescence screen, and a YAG ($Y_3Al_5O_{12}$) crystal. The fluorescence screen, which is mounted as a powder-coated vacuum viewport on the vacuum setup, is used to view the diffraction patterns for the corresponding photon energies. Figure~\ref{fig:spot}a) displays, from top to bottom, the results obtained from the SiC/Si mirror in combination with an Al filter, the grating, and the grating in combination with the Al filter, respectively. For these measurements, the zeroth order diffraction spot from the grating is aligned to the same position on the fluorescence window as the spot from the mirror. It is clearly seen that the grating generates resolvable diffraction orders with spatial separation. For the bottom panel of Fig.\ref{fig:spot}a), the spots on the far left and far right are blocked by the Al filter. This verifies that the remaining spots are indeed the fifth harmonic ($h\nu$ = 18.1~eV) spots from the HHG, as the Al filter provides a band-pass filter for photon energies above $\sim$15~eV, thus blocking the third harmonic light ($\sim$10.8~eV). The spatial separation between the first order spots for 10.8~eV and 18.1~eV, obtained from the image, is about 7~mm. The experimental result is consistent with the simulation presented in the inset of Fig.\ref{fig:sim}a). The 7~mm distance is sufficiently large so that the two harmonics can be selectively filtered out using a small aperture, in case one would like to prevent the unwanted higher orders from reaching the sample. The latter is not always necessary since the distance puts the unwanted orders outside the view of the electron analyzer of the photoemission setup.. In the middle panel of Fig.\ref{fig:spot}a), the very bright zero order spot, located in the center of the image, is blocked by a metallic foil to avoid damaging the fluorescence window and to allow the weaker first order spots to be imaged with the camera.

 A YAG crystal is used for accurate determination of the light spot size produced by the two NIM setups. The YAG crystal is placed at the sample position in the photoemission vacuum chamber, and its exact position can be controlled by a motorized manipulator (SPECS) with a calibrated millimeter scale. Figure~\ref{fig:spot}b) and c) show the results of the fifth harmonic ($h\nu$ = 18.1~eV), as imaged by the spherical SiC/Si mirror and the grating, respectively. The spherical mirror provides a spot size of about 104$\times$95~{\textmu}m$^2$ (H$\times$V), while the grating gives 123$\times$144~{\textmu}m$^2$. From ray-tracing simulations, presuming a beam waist of the HHG radiation of 10~{\textmu}m and a divergence of 6~mrad, we obtain a focus of 47.7$\times$48.1~{\textmu}m$^2$ (H$\times$V), $\sim$2.1 and $\sim$2.7 times smaller than the measured values, respectively. The deviation between simulated and measured spot sizes for the mirror+filter configuration, is possibly due to underestimation of the HHG plasma size, or the experimental spherical aberration being larger than the ones used in the simulation. The 10~{\textmu}m HHG source size, determined by the focus of the off-axis parabolic mirror ($f$ = 101.6~mm), is in practice difficult to realize due to several reasons, including misalignment, UV beam quality, HHG phase-matching, as well as the roughness of a mirror coating. These uncertainties are present for both types of NIMs, but the additional broadening of the spot produced by the spherical grating mainly comes from the fabrication imperfections, as discussed previously.

\subsection{Efficiency}
\begin{figure}[htp]
\centering
\includegraphics[width=\textwidth]{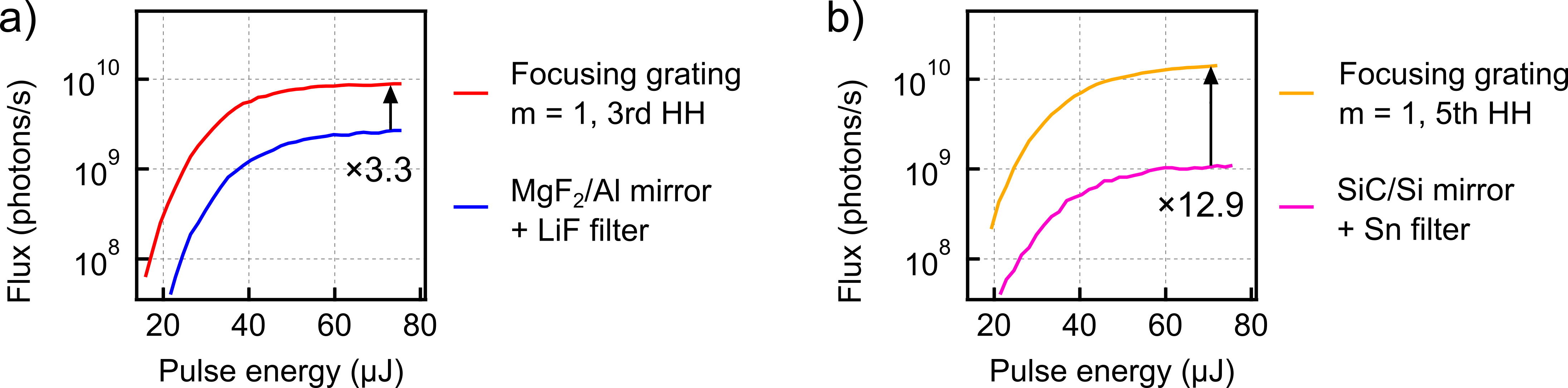}
\caption{\label{fig:flux}Flux measurement of the mirror+filter and focusing grating NIM for a) 10.8 eV and b)18.1 eV photon energy. The results are plotted as a function of the pulse energy driving the HHG. The specific configuration of the monochromator, as well as the efficiency enhancement are specified in the figure.}
\end{figure}

Figure~\ref{fig:flux} shows the photon flux measurement for the two types of monochromators at 10.8~eV and 18.1~eV photon energies, reflecting the efficiency. The photon flux is determined by measuring the drain current from a piece of annealed tantalum foil placed in the beam path. The yield efficiency is estimated based on the Ref. \citeonline{feuerbacher1972experimental}. The laser that drives the HHG operates at a repetition rate of 250~kHz, and the pulse energy is tunable up to approximately 80~{\textmu}J. The flux measurement using the focusing grating is based on the first order diffraction spots from the corresponding harmonics. Overall, the focusing grating provides a considerable enhancement of the monochromator efficiency compared to the mirror+filter solution. In the case of our design that is specifically optimized for the 18.1~eV photon energy, there is an order of magnitude improvement in the photon flux compared the mirror+filter based solution.

\subsection{Temporal broadening}
\begin{figure}[htp]
\centering
\includegraphics[width=\textwidth]{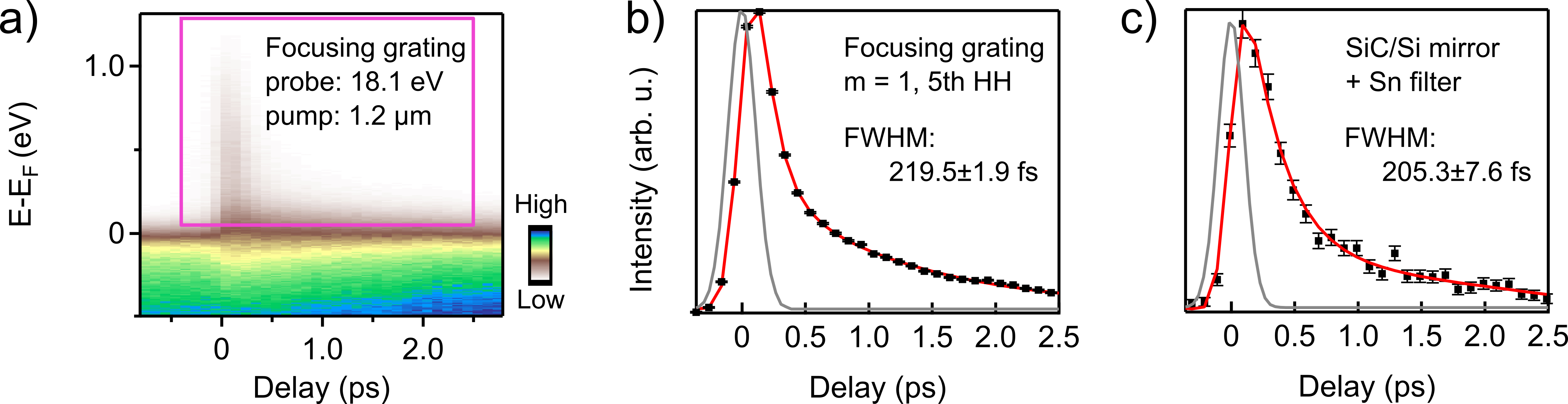}
\caption{\label{fig:graphene}Time- and angle-resolved photoemission spectroscopy measurements on a monolayer p-doped graphene, performed with 18.1 eV (third harmonic). a) The momentum-integrated spectrum as a function of time delay, taken by using focusing grating NIM. The pump beam wavelength is 1.2~{\textmu}m. Purple box indicates the energy integration window used to obtain data shown in b) and c). b) The excitation intensity as a function of delay (black dots) and fit to the data (red line). The gray line shows the fitted Gaussian peak representing the temporal resolution. c) Same as in b), but for the mirror+filter NIM. FWHM stands for full width at half maximum. All data taken at room temperature.}
\end{figure}

Figure~\ref{fig:graphene} compares pump-probe measurements of the electronic excitation in graphene measured with the tr-ARPES setup described in Ref.~\citeonline{guo2022narrow}, and using the two different types of monochromatization. Since the fifth harmonic with 18.1~eV photon energy has the shortest pulse duration of the two wavelengths considered here\cite{guo2022narrow}, the temporal properties were investigated using this photon energy. A p-type graphene sample was selected as a test specimen since it has sufficiently fast intrinsic electron dynamics to precisely reflect the system-limited temporal resolution\cite{gierz2014non,johannsen2013direct}. The sample used is a quasi-freestanding monolayer graphene on a 6H-SiC (0001) substrate\cite{forti2011large}. The measurement was conducted at room temperature, and the data acquisition time for each delay point is 5~min.

Figure~\ref{fig:graphene}a) shows the excitation spectra with integrated momentum as a function of delay time, taken with the focusing grating NIM. The pump wavelength is 1.2~{\textmu}m, and the fluence is approximately 35~{\textmu}J/cm$^2$. In Fig.~\ref{fig:graphene}b) the energy-integrated intensity within the purple box drawn in Fig.~\ref{fig:graphene}a) is plotted as dot markers. The fit (red line) to the data points is a convolution of a 2-$\tau$-parameter exponential decay curve and a Gaussian function, using the two $\tau$'s, the decay curve amplitude, and the Gaussian standard deviation $\sigma$ as fitting parameters. The full width at half maximum (FWHM) of the fitted Gaussian distribution (gray line) represents the overall temporal resolution of the system, while the $\tau$-parameters describe the rate of decay after the excitation. Similar to Fig.~\ref{fig:graphene}b), the data in Fig.~\ref{fig:graphene}c) is acquired by using the mirror+filter solution. The same energy window for integration is selected as for the grating case. 

For the focusing grating NIM, an overall temporal resolution of about 219.5$\pm$1.9~fs is found, while the mirror+filter results in a 205.3$\pm$7.6~fs temporal resolution. These results suggest that the focusing grating NIM introduces a temporal broadening of the probe pulse, as expected. By deconvoluting the broadening from the pump beam, the pulse duration of the probe beam is found to be $\sim$195~fs and $\sim$179~fs, for the focusing grating and mirror+filter, respectively. In comparison, the contribution from the grating monochromator is about 72.4$\pm$27.4~fs, which agrees well with the calculation presented in Fig.~\ref{fig:sim}b). As the additional broadening from the grating based NIM is only 14~fs (increase by 6.8\%) compared to the mirror+filter, this shows that the influence of the grating on the temporal resolution is negligible in many circumstances. The two sets of data were collected with the same pulse energy used to drive the HHG ($\sim$60{\textmu}J), and the same acquisition time. The statistics presented here, as indicated by the noise level and error bars, reflects the increased photon flux at 18.1~eV photon energy when using the focusing grating (error bars for grating data is smaller than the measurement points).

\section{Discussion and Conclusion}
Figure~\ref{fig:flux} shows that using a low-density focusing grating for monochromatization of 10.8~eV and 18.1~eV photon energies from a HHG source provides higher efficiency compared to a solution based on a focusing mirror in combination with thin film filters. With the grating, a temporal broadening of $<$10\% is found for laser pulses with a duration on the order of 200~fs. Since the particular HHG setup used in this work is designed to achieve a high energy resolution, and thus have relatively long laser pulses, the pulse duration of the driving laser is 460~fs~\cite{guo2022narrow}, and the pulse stretching caused by the grating has little influence on the overall temporal resolution of the system. For HHG sources with shorter pulses, however, the pulse-front tilt induced by the focusing grating can be minimized by reducing the line density. However, this has to be weighted against the reduction in resolving power of the grating, as well as the spatial separation of the diffraction orders at the sample position. In the present work, a line density of 40 lines/mm was selected since this gave the best trade-off between the temporal broadening and the separation of the harmonics. The pulse-front tilt can also be reduced by decreasing the beam size on the grating by an iris or an aperture. Doing so, however, comes at the cost of reduced photon flux. Since the grating based solution provides considerably higher transmission this is however a viable trade-off strategy that provides increased flexibility. 

It is worth noting that very few thin film filters work well in the low photon energy range below 20 eV due to high absorption at these wavelengths. In practice, band-pass thin film filters normally require ultra-thin thickness to have decent transmission. Some materials, such as Sn or Si filters ($\sim$ 100~nm), are very delicate and can easily be destroyed by handling or by absorption of residual light from the driving laser. Furthermore, frequent replacement is time-consuming and costly. As a filter-less solution, the focusing grating circumvents these challenges. Applying the focusing grating in the higher photon energy range above 20~eV could be achieved by optimizing the groove depth, in combination with a multi-layer coating specifically tailored for the photon energy of interest. In terms of the fabrication quality of the grating, the most pressing issue is the curvature of the surface, as that can cause a non-uniformity of the etched pattern. This can be addressed by further optimizing the specific procedures, in particular the lithography process, or by using an error-compensated photo-mask.

In summary, we have compared a monochromator design based on a normal-incidence spherical grating and compared it to a solution based on a spherical mirror and filter. The experimental comparison targets the characteristics important for applications in ultrafast electron spectroscopy. We have designed and fabricated a spherical grating as a compact, single-element monochromator, specifically optimized for a photon energy of 18.1~eV. The performance of the focusing grating monochromator is compared to a solution based on a spherical mirror in combination with a thin film filter -- the latter is a commonly used monochromatization scheme for HHG-based ultrafast light sources. Overall, the mirror+filter solution can provide a smaller spot size at the focus with no temporal broadening. The focusing grating can, on the other hand, provide an order of magnitude higher efficiency in terms of photon flux with only a small amount of focus size increase and temporal broadening. The single-element design of the grating enables a very compact optical layout with minimal need for maintenance. The focusing grating, furthermore, requires only a low-cost, commercially available Si-mirror substrate, and can be produced using standard nano-fabrication processing. The final product is both mechanically robust, and can handle a high power loads. In short, the mirror+filter solution is more appropriate in cases where there is no tolerance in temporal broadening, while the focusing grating is more advantageous in efficiency. The presented grating solution also presents significant advantages in terms of low-cost, compact size and high efficiency, and could be relevant for applications outside of photoelectron spectroscopy as well. 

\begin{backmatter}
\bmsection{Funding}
This work was financially supported by the Knut and Alice Wallenberg foundation (No.2018-0104) and the Swedish research council VR (No.2019-00701). Q.G acknowledges the fellowship from China scholarship council (No.201907930007). M.D. acknowledges financial support from the Göran Gustafsson foundation. 

\bmsection{Acknowledgments}
We acknowledge the Albanova nanolab for the device fabrication. We thank Dr. Bharti Matta and professor Ulrich Starke for providing the high-quality graphene sample.

\bmsection{Disclosures}
The authors declare no conflicts of interest.

\bmsection{Data Availability Statement}
The data that support the findings of this study are available
from the corresponding author upon reasonable request.

\bmsection{Supplemental document}
See Supplement 1 for supporting content. 

\end{backmatter}



\bibliography{1_maintexts}

\begin{thebibliography}{10}
\newcommand{\enquote}[1]{``#1''}

\bibitem{LaserHHG_KepteynMurnanePRL1996}
J.~Zhou, J.~Peatross, M.~M. Murnane, H.~C. Kapteyn, and I.~P. Christov,
  \enquote{Enhanced high-harmonic generation using 25 fs laser pulses,}
  {\protect\JournalTitle{Phys. Rev. Lett.}} \textbf{76}, 752--755 (1996).

\bibitem{HHG_CoherentSoftXrays_Murnane_PRL1997}
Z.~Chang, A.~Rundquist, H.~Wang, M.~M. Murnane, and H.~C. Kapteyn,
  \enquote{Generation of coherent soft x rays at 2.7 nm using high harmonics,}
  {\protect\JournalTitle{Phys. Rev. Lett.}} \textbf{79}, 2967--2970 (1997).

\bibitem{HHG_CoherentXrays_WaterWindow_Krausz_Science1997}
C.~Spielmann, N.~H. Burnett, S.~Sartania, R.~Koppitsch, M.~Schnürer, C.~Kan,
  M.~Lenzner, P.~Wobrauschek, and F.~Krausz, \enquote{Generation of coherent
  {X-rays} in the water window using 5-femtosecond laser pulses,}
  {\protect\JournalTitle{Science}} \textbf{278}, 661--664 (1997).

\bibitem{HHG_Ne_Ar_LHullier_PRA1993}
C.-G. Wahlstr\"om, J.~Larsson, A.~Persson, T.~Starczewski, S.~Svanberg,
  P.~Sali\`eres, P.~Balcou, and A.~L'Huillier, \enquote{High-order harmonic
  generation in rare gases with an intense short-pulse laser,}
  {\protect\JournalTitle{Phys. Rev. A}} \textbf{48}, 4709--4720 (1993).

\bibitem{HHG_threeStepModel}
R.~Santra and A.~Gordon, \enquote{Three-step model for high-harmonic generation
  in many-electron systems,} {\protect\JournalTitle{Phys. Rev. Lett.}}
  \textbf{96}, 073906 (2006).

\bibitem{tate2007scaling}
J.~Tate, T.~Auguste, H.~Muller, P.~Sali{\`e}res, P.~Agostini, and L.~DiMauro,
  \enquote{Scaling of wave-packet dynamics in an intense midinfrared field,}
  {\protect\JournalTitle{Physical Review Letters}} \textbf{98}, 013901 (2007).

\bibitem{comby2019cascaded}
A.~Comby, D.~Descamps, S.~Beauvarlet, A.~Gonzalez, F.~Guichard, S.~Petit,
  Y.~Zaouter, and Y.~Mairesse, \enquote{Cascaded harmonic generation from a
  fiber laser: a milliwatt {XUV} source,} {\protect\JournalTitle{Optics
  Express}} \textbf{27}, 20383--20396 (2019).

\bibitem{Mono_CzernyTurner1930}
M.~Czerny and A.~F. Turner, \enquote{Über den astigmatismus bei
  spiegelspektrometern,} {\protect\JournalTitle{Zeitschrift f{\"u}r Physik}}
  \textbf{61}, 792 -- 797 (1930).

\bibitem{peatman1995exactly}
W.~Peatman, J.~Bahrdt, F.~Eggenstein, G.~Reichardt, and F.~Senf, \enquote{The
  exactly focusing spherical grating monochromator for undulator radiation at
  bessy,} {\protect\JournalTitle{Review of Scientific Instruments}}
  \textbf{66}, 2801--2806 (1995).

\bibitem{borisenko2012one}
S.~Borisenko, \enquote{“{One}-cubed” arpes user facility at {BESSY II},}
  {\protect\JournalTitle{Synchrotron Radiation News}} \textbf{25}, 6--11
  (2012).

\bibitem{hoffmann2004undulator}
S.~Hoffmann, C.~S{\o}ndergaard, C.~Schultz, Z.~Li, and P.~Hofmann, \enquote{An
  undulator-based spherical grating monochromator beamline for angle-resolved
  photoemission spectroscopy,} {\protect\JournalTitle{Nuclear Instruments and
  Methods in Physics Research Section A: Accelerators, Spectrometers, Detectors
  and Associated Equipment}} \textbf{523}, 441--453 (2004).

\bibitem{poletto2004time}
L.~Poletto, \enquote{Time-compensated grazing-incidence monochromator for
  extreme-ultraviolet and soft {X-ray} high-order harmonics,}
  {\protect\JournalTitle{Applied Physics B}} \textbf{78}, 1013--1016 (2004).

\bibitem{poletto2009time}
L.~Poletto, P.~Villoresi, F.~Frassetto, F.~Calegari, F.~Ferrari, M.~Lucchini,
  G.~Sansone, and M.~Nisoli, \enquote{Time-delay compensated monochromator for
  the spectral selection of extreme-ultraviolet high-order laser harmonics,}
  {\protect\JournalTitle{Review of Scientific Instruments}} \textbf{80}, 123109
  (2009).

\bibitem{ito2010spatiotemporal}
M.~Ito, Y.~Kataoka, T.~Okamoto, M.~Yamashita, and T.~Sekikawa,
  \enquote{Spatiotemporal characterization of single-order high harmonic pulses
  from time-compensated toroidal-grating monochromator,}
  {\protect\JournalTitle{Optics Express}} \textbf{18}, 6071--6078 (2010).

\bibitem{frassetto2011single}
F.~Frassetto, C.~Cacho, C.~A. Froud, I.~E. Turcu, P.~Villoresi, W.~A. Bryan,
  E.~Springate, and L.~Poletto, \enquote{Single-grating monochromator for
  extreme-ultraviolet ultrashort pulses,} {\protect\JournalTitle{Optics
  Express}} \textbf{19}, 19169--19181 (2011).

\bibitem{igarashi2012pulse}
H.~Igarashi, A.~Makida, M.~Ito, and T.~Sekikawa, \enquote{Pulse compression of
  phase-matched high harmonic pulses from a time-delay compensated
  monochromator,} {\protect\JournalTitle{Optics Express}} \textbf{20},
  3725--3732 (2012).

\bibitem{grazioli2014citius}
C.~Grazioli, C.~Callegari, A.~Ciavardini, M.~Coreno, F.~Frassetto, D.~Gauthier,
  D.~Golob, R.~Ivanov, A.~Kivim{\"a}ki, B.~Mahieu \emph{et~al.},
  \enquote{Citius: An infrared-extreme ultraviolet light source for fundamental
  and applied ultrafast science,} {\protect\JournalTitle{Review of Scientific
  Instruments}} \textbf{85}, 023104 (2014).

\bibitem{ojeda2016harmonium}
J.~Ojeda, C.~Arrell, J.~Grilj, F.~Frassetto, L.~Mewes, H.~Zhang, F.~Van~Mourik,
  L.~Poletto, and M.~Chergui, \enquote{Harmonium: A pulse preserving source of
  monochromatic extreme ultraviolet (30--110 ev) radiation for ultrafast
  photoelectron spectroscopy of liquids,} {\protect\JournalTitle{Structural
  Dynamics}} \textbf{3}, 023602 (2016).

\bibitem{heimann2011linac}
P.~Heimann, O.~Krupin, W.~F. Schlotter, J.~Turner, J.~Krzywinski,
  F.~Sorgenfrei, M.~Messerschmidt, D.~Bernstein, J.~Chalupsk{\`y},
  V.~H{\'a}jkov{\'a} \emph{et~al.}, \enquote{Linac coherent light source soft
  {X}-ray materials science instrument optical design and monochromator
  commissioning,} {\protect\JournalTitle{Review of Scientific Instruments}}
  \textbf{82}, 093104 (2011).

\bibitem{frassetto2014time}
F.~Frassetto, E.~Ploenjes, M.~Kuhlmann, and L.~Poletto,
  \enquote{Time-delay-compensated grating monochromator for fel beamlines,} in
  \emph{{X-Ray} Free-Electron Lasers: Beam Diagnostics, Beamline
  Instrumentation, and Applications II,}  vol. 9210 (International Society for
  Optics and Photonics, 2014), p. 92100I.

\bibitem{fabris2019high}
N.~Fabris, P.~Miotti, F.~Frassetto, and L.~Poletto, \enquote{A high resolution
  {XUV} grating monochromator for the spectral selection of ultrashort harmonic
  pulses,} {\protect\JournalTitle{Applied Sciences}} \textbf{9}, 2502 (2019).

\bibitem{poletto2014double}
L.~Poletto, P.~Miotti, F.~Frassetto, C.~Spezzani, C.~Grazioli, M.~Coreno,
  B.~Ressel, D.~Gauthier, R.~Ivanov, A.~Ciavardini \emph{et~al.},
  \enquote{Double-configuration grating monochromator for extreme-ultraviolet
  ultrafast pulses,} {\protect\JournalTitle{Applied Optics}} \textbf{53},
  5879--5888 (2014).

\bibitem{frassetto2014grating}
F.~Frassetto, P.~Miotti, and L.~Poletto, \enquote{Grating configurations for
  the spectral selection of coherent ultrashort pulses in the
  extreme-ultraviolet,} in \emph{Photonics,}  vol.~1 (Multidisciplinary Digital
  Publishing Institute, 2014), pp. 442--454.

\bibitem{guo2022narrow}
Q.~Guo, M.~Dendzik, A.~Grubi{\v{s}}i{\'c}-{\v{C}}abo, M.~H. Berntsen, C.~Li,
  W.~Chen, B.~Matta, U.~Starke, B.~Hessmo, J.~Weissenrieder \emph{et~al.},
  \enquote{A narrow bandwidth extreme ultra-violet light source for time-and
  angle-resolved photoemission spectroscopy,} {\protect\JournalTitle{Structural
  Dynamics}} \textbf{9}, 024304 (2022).

\bibitem{feuerbacher1972experimental}
B.~Feuerbacher and B.~Fitton, \enquote{Experimental investigation of
  photoemission from satellite surface materials,}
  {\protect\JournalTitle{Journal of Applied Physics}} \textbf{43}, 1563--1572
  (1972).

\bibitem{gierz2014non}
I.~Gierz, S.~Link, U.~Starke, and A.~Cavalleri, \enquote{Non-equilibrium dirac
  carrier dynamics in graphene investigated with time-and angle-resolved
  photoemission spectroscopy,} {\protect\JournalTitle{Faraday Discussions}}
  \textbf{171}, 311--321 (2014).

\bibitem{johannsen2013direct}
J.~C. Johannsen, S.~Ulstrup, F.~Cilento, A.~Crepaldi, M.~Zacchigna, C.~Cacho,
  I.~E. Turcu, E.~Springate, F.~Fromm, C.~Raidel \emph{et~al.}, \enquote{Direct
  view of hot carrier dynamics in graphene,} {\protect\JournalTitle{Physical
  Review Letters}} \textbf{111}, 027403 (2013).

\bibitem{forti2011large}
S.~Forti, K.~Emtsev, C.~Coletti, A.~Zakharov, C.~Riedl, and U.~Starke,
  \enquote{Large-area homogeneous quasifree standing epitaxial graphene on
  {SiC} (0001): Electronic and structural characterization,}
  {\protect\JournalTitle{Physical Review B}} \textbf{84}, 125449 (2011).

\end{thebibliography}






\end{document}


\centerline{\large{Supplementary Material for}}
\title{Experimental comparison of normal-incidence grating and filter-based monochromatization schemes for time- and angle-resolved photoemission spectroscopy}

\author{Qinda Guo$^{1}$, Maciej Dendzik$^{1}$, Magnus H. Berntsen$^{1}$, Antonija  Grubi\v{s}i\'{c}-\v{C}abo$^{1,2}$, Cong Li$^{1}$, Wanyu Chen$^{1}$, Yang Wang$^{1}$ and Oscar Tjernberg$^{1,*}$
}

\affiliation{
\\$^{1}$Department of Applied Physics, KTH Royal Institute of Technology, Hannes Alfv\'{e}ns v\"{a}g 12, 114 19 Stockholm, Sweden
\\$^{2}$Zernike Institute for Advanced Materials, University of Groningen, 9747 AG Groningen, The Netherlands
\\$^{*}$Electronic mail: oscar@kth.se
}

\pacs{}

\maketitle

{\noindent \bf 1. Dependence of the focusing grating efficiency on the coating thickness}

\begin{figure}[htbp]
\centering
\includegraphics[scale=0.078]{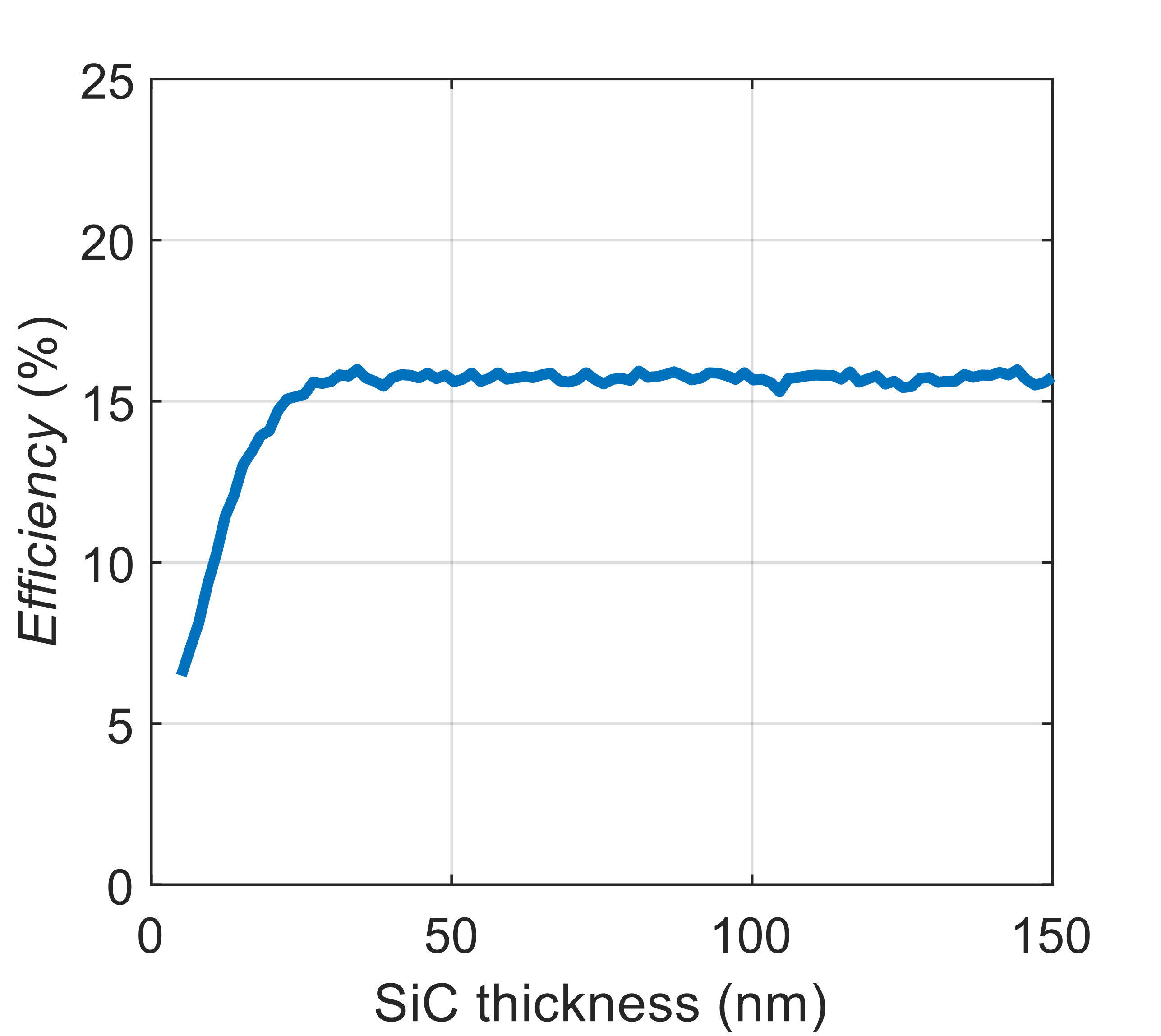}
\caption{\label{fig:coating} Overall mono efficiency dependence on the coating thickness.}
\end{figure}

\begin{figure}[htbp]
\centering
\includegraphics[scale=0.078]{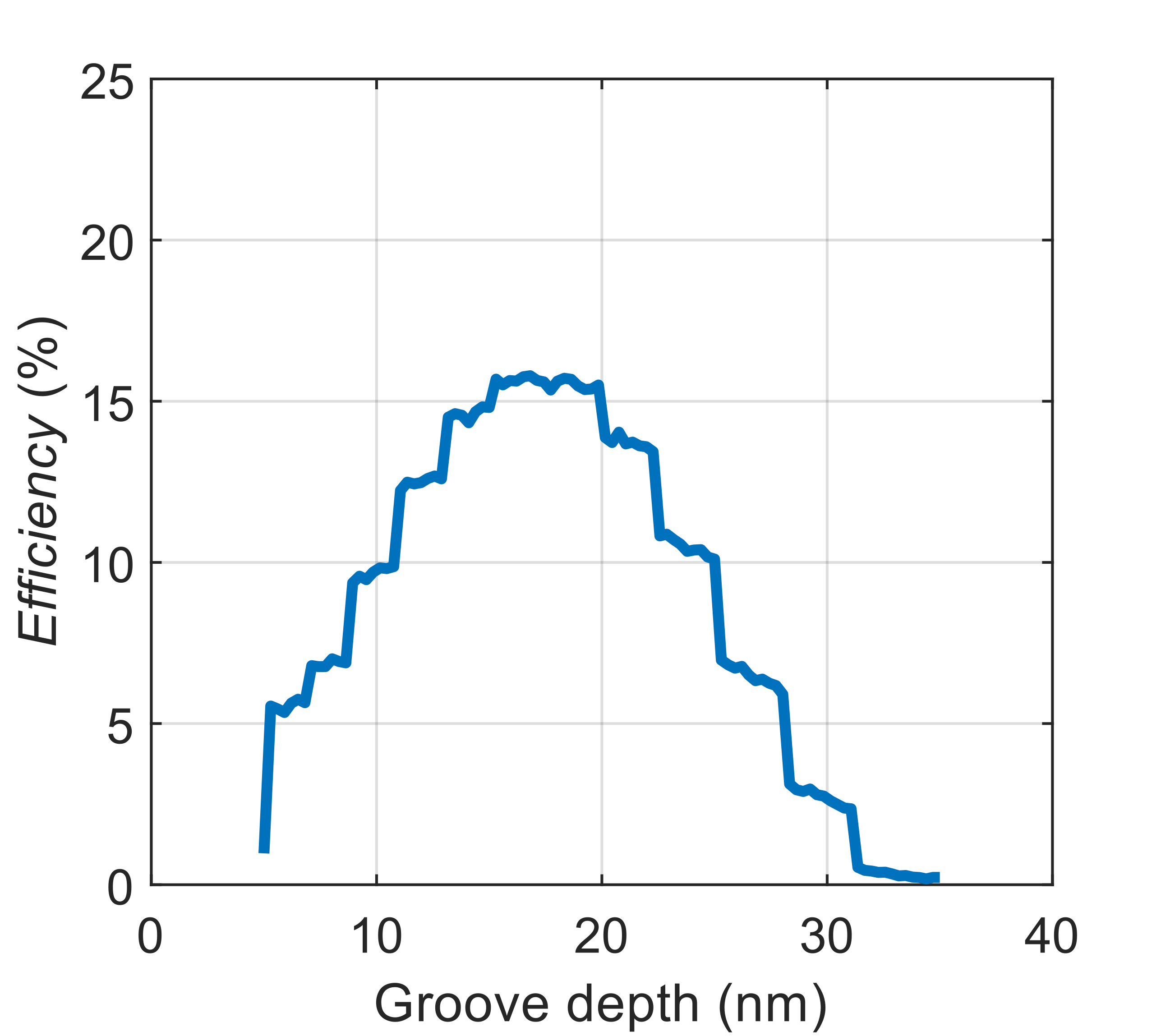}
\caption{\label{fig:groove} Overall mono efficiency dependence on the groove depth.}
\end{figure}

\begin{figure}[htbp]
\centering
\includegraphics[scale=0.078]{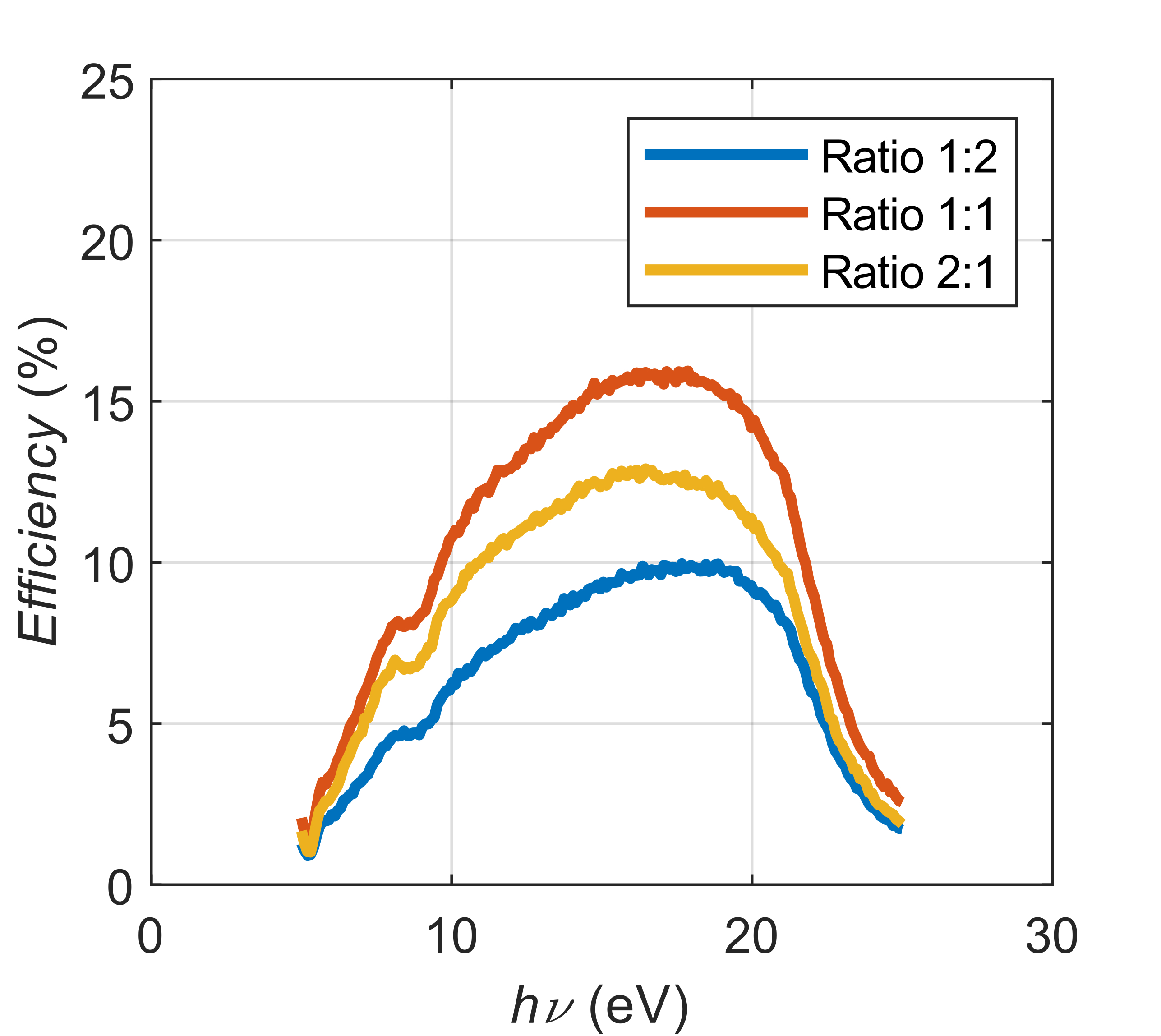}
\caption{\label{fig:ratios} Overall mono efficiency dependence on the groove-to-land ratio.}
\end{figure}

\noindent Figure~\ref{fig:coating} shows the calculated overall monochromator efficiency for the focusing grating as a function of the SiC coating thickness. Note that the results demonstrated in Fig.~\ref{fig:coating}, Fig.~\ref{fig:groove} and Fig.~\ref{fig:ratios} are based on tuning the single-variable, while the rest of the parameters are fixed, as elaborated in the main texts. The calculation method is based on Ref.~\onlinecite{baumgartel2016ray}. Figure~\ref{fig:coating} shows that the efficiency stays almost constant above coating thickness for approximately 30~nm. The experimentally characterized grating has coating thickness of about 80~nm.

{\noindent \bf 2. Dependence of the focusing grating efficiency on the groove depth}

\noindent Figure~\ref{fig:groove} shows the result of efficiency as a function of groove depth. It shows that the efficiency of the grating is almost unchanged for grooves between 15~nm and to 20~nm in depth. The fabricated piece has an etched depth of about 18.4$\pm$2.5~nm, which is within the best performance range, based on the calculation.

{\noindent \bf 3. Dependence of the focusing grating efficiency on the groove-to-land ratio}

\noindent Figure~\ref{fig:ratios} shows the influence of the groove-to-land ratio parameter on the monochromator efficiency. It is clearly seen that the efficiency decreases for ratios either lower or higher than 1. This is a sensitive parameter affecting the performance, as indicated by the calculation. The fabricated device has a ratio smaller than 1, which might cause poorer performance than expected.

{\noindent \bf 4. The simulation of the transmission window for the thin film filters}

\begin{figure}[htbp]
\centering
\includegraphics[scale=0.078]{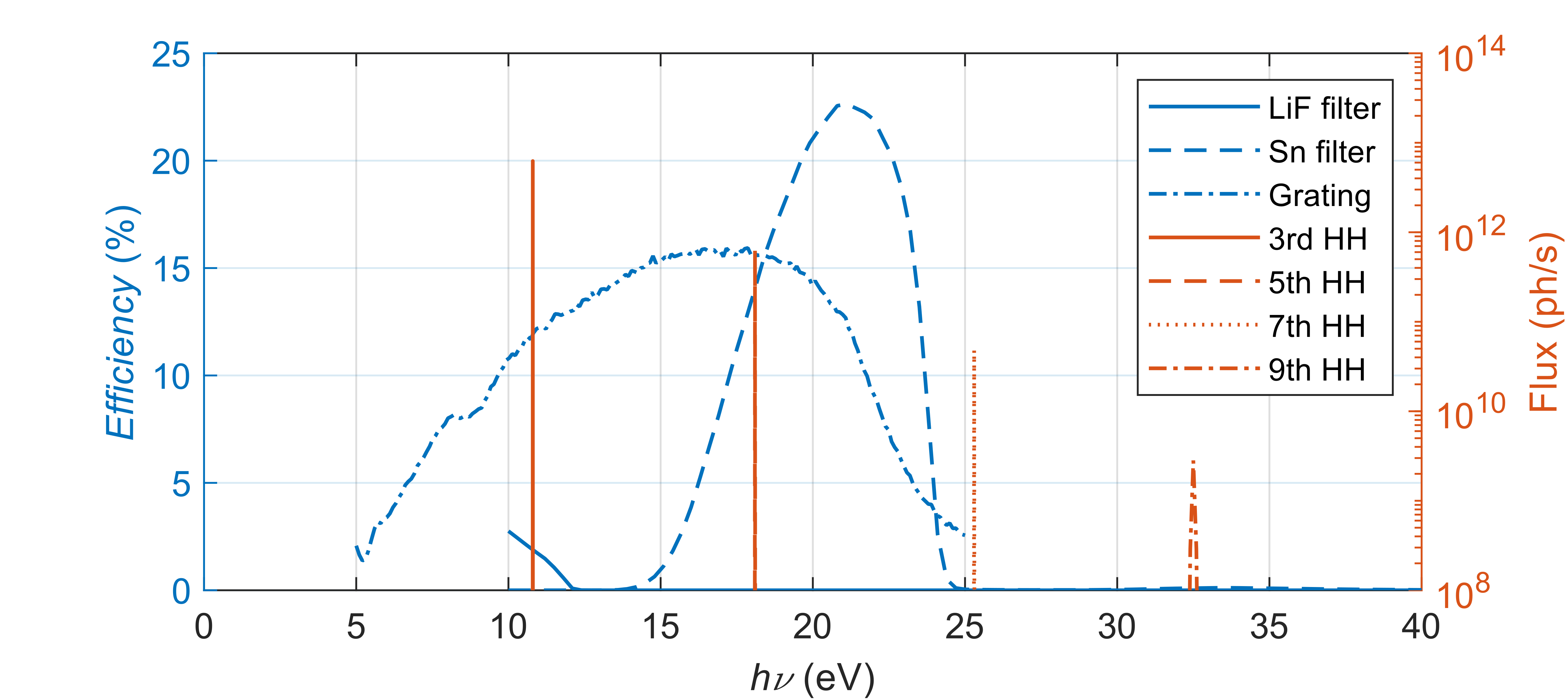}
\caption{\label{fig:filters} The calculated transmission window of the thin film filters - LiF and Sn, as a function of photon energies up till 40~eV. To compare in parallel, the spherical grating efficiency as the function of photon energy, as well as the harmonics spectrum of HHG, are plotted. The blue-color and red-color plot correspond to the left and right y-axis, respectively.}
\end{figure}

\noindent Figure~\ref{fig:filters} shows the calculation to the transmission window of the thin film filters employed in the setup \cite{guo2022narrow}, according to Ref.~\onlinecite{henke1993x}. The spherical grating efficiency and the HHG spectrum are also plotted for ease of comparison.

\vspace{3mm}


\noindent\small[1]	\;P. Baumgärtel, M. Witt, J. Baensch, M. Fabarius, A. Erko, F. Schäfers, and H. Schirmacher, in \textit{AIP}\\ \indent \; \textit{Conference Proceedings}, Vol. 1741 (AIP Publishing LLC, 2016) p. 040016.\\
\noindent\small[2]	\;Q. Guo, M. Dendzik, A. Grubi\v{s}i\'{c}-\v{C}abo, M. H. Berntsen, C. Li, W. Chen, B. Matta, U. Starke, B. Hes-
\indent \; smo, J. Weissenrieder, \textit{et al}., Structural Dynamics 9, 024304 (2022).\\
\noindent\small[3]	\; B. L. Henke, E. M. Gullikson, and J. C. Davis, Atomic data and nuclear data tables \textbf{54}, 181 (1993).\\

\newpage